\documentclass[twocolumn,prb,aps,epsf,amsmath,amssymb,showpacs]{revtex4}
\usepackage{comment}
\usepackage[dvipsnames]{xcolor}

\usepackage{yfonts}

\usepackage{graphicx}
\usepackage{dcolumn}
\usepackage{bm}

\newcommand{\nn}{\nonumber}

\def\bea{\begin{equation}\begin{aligned}}
\def\eea{\end{aligned}\end{equation}}

\begin{document}

\title{Fermi Liquid Theory for Spin Current of a Ferromagnet}
\author{Wayne M. Saslow}
\email{wsaslow@tamu.edu}
\affiliation{ Texas A\&M University, College Station, Texas, 77843, U.S.A.}
\author{Chen Sun}
\email{chensun@hnu.edu.cn}
\affiliation{ School of Physics and Electronics, Hunan University, Changsha 410082, China}
\begin{abstract}
A recent work [arXiv:2402.04639] considered the dynamical equations for ferromagnets using Onsager's irreversible thermodynamics with fundamental variables magnetization $\vec{M}$ and spin current $\vec{J}_{i}$. The resulting equations have the same structure as Leggett's Fermi liquid theory for the nuclear paramagnet $^{3}$He. Specifically, $\partial_{t}\vec{J}_{i}$ contains a term varying as $\partial_{i}\vec{M}$ that we interpret as associated with a vector spin pressure, and a term giving a mean-field along $\vec{M}$, about which $\vec{J}_{i}$ precesses.   (There is also a slow decay term in $\partial_{t}\vec{M}$ not normally present in the Leggett equations, which are intended for shorter-time spin-echo experiments.)
The present work applies Fermi liquid theory to $\vec{J}_{i}$ of ferromagnets. The resulting dynamical equation for $\vec J_i$ confirms the form of $\vec J_i$ found in [arXiv:2402.04639], but now the previously unknown non-dissipative parameters are given in terms of the quasiparticle interaction parameters of Fermi liquid theory. 
In the paramagnetic limit the  present theory agrees with Leggett and related work.

\end{abstract}

\date{\today}

\maketitle

\section{Introduction}


Spintronics is an area of great contemporary interest.\cite{Wolf05,ZFD-RMP,Chumak15,Hirohata20,Kim24} In order to control and manipulate spins or spin currents, it is important to know how they evolve with time.  
Therefore, theoretical understanding of spin dynamics in magnetic systems is a central question in spintronics.

Onsager's irreversible thermodynamics has been a useful tool for studying the dynamics of various magnetic systems, including paramagnets, ferromagnets and spin glasses.\cite{JohnsonSilsbee87,Saslow07,Saslow15,Saslow17,SaslowSunXu22,SunSaslow23,SunSaslow24}  However, the dynamical equations of irreversible thermodynamics do not give the values of the parameters it contains, which must be determined by more microscopic theories.  The dissipative terms, typically involving either decay or diffusion, are especially difficult to obtain, and are not considered in the present work. 

Using Onsager's irreversible thermodynamics, a recent work\cite{SunSaslow24} considered the spin dynamics for ferromagnets with fundamental variables magnetization $\vec M$ and spin current $\vec J_i$.  It gave dynamical equations for $\vec M$ and $\vec{J}_{i}$:
\begin{equation}
\partial_{t}\delta\vec{M}+\partial_{i}\vec{J}_{i}=-\gamma\delta\vec{M}\times\vec{B}-\frac{1}{\tau_{M}}\delta\vec{M},
\label{dM/dt-time}
\end{equation}
\begin{equation}
\partial_{t}\vec{J}_{i}+G\partial_{i}\delta\vec{M}=-\gamma\vec{J}_{i}\times(\vec{B}+\lambda\vec{M})
-\frac{1}{\tau_{\vec{J}}}\vec{J}_{i},
\label{dJM/dt-time}
\end{equation}
where in the second equation  for $\partial_t\vec{J}_{i}$  the two constants $G$ and $\lambda$ are unknown.  ($G$ has dimensions of velocity squared and $\lambda$ has, in SI units, dimensions of $\mu_{0}$.)  In a nuclear spin system like $^{3}$He, $\tau_{M}^{-1}$ is very small, so for most purposes magnetization decay can be neglected, and the above equations have the same form as the Leggett equations for nuclear paramagnets.\cite{Leggett70} For electronic ferromagnets, $\tau_{M}^{-1}$ is essential, and $\vec{M}$ is spontaneous.  Thus magnets satisfy a slightly modified version of the Leggett equations.

The present work applies Fermi liquid theory to the $\vec{J}_{i}$ of ferromagnets; it corresponds to the Onsager theory for a ferromagnetic system presented in [\onlinecite{SunSaslow24}]. Specifically, it derives a dynamical equation for the spin current $\vec J_i$, which has the same the form as in [\onlinecite{SunSaslow24}], but now with the previously unknown non-dissipative parameters determined by the quasiparticle interaction parameters of Fermi liquid theory.

Landau's original works on Fermi liquid theory are [\onlinecite{LandauFLT56}], which considers the magnetic susceptibility, and [\onlinecite{LandauFLT57}], which considers spin-dependent zero sound.  They largely treat the 2-by-2 spin-space number and energy operators as diagonal.  
Following Landau, Silin wrote two papers [\onlinecite{Silin2}] and [\onlinecite{Silin1}] which explicitly discuss the effects of spin in a paramagnetic Fermi liquid. 
Silin obtained a kinetic equation for the distribution function, Eq.~(1.8) in [\onlinecite{Silin1}]. Starting from Silin's kinetic equation, Leggett \cite{Leggett70} considered explicit forms for the Fermi liquid interactions between quasiparticles and found a dynamical equation for the spin current with parameters given in terms of the interaction parameters. (See Eq. (20) in Ref.~[\onlinecite{Leggett70}].) Relative to a paramagnet, the spin current in a ferromagnet is complicated by having two Fermi surfaces. Nevertheless, the Fermi liquid theory of paramagnets will be useful, both because some of its definitions can be directly used for ferromagnets, and because it serves as a limit for the Fermi liquid theory of ferromagnets. Additional theoretical works on spin-polarized Fermi systems -- solids, liquids, and gases -- include [\onlinecite{Mineev04,Mineev05,Bedell86,JeonMullin89,MiyakeMullinStamp85,GolosovRuckenstein95,Levitov02}].

This paper is organized as follows. As background, Sect.~\ref{s:density} defines density and polarization density. Sect.~\ref{s:LFLT} presents the basics of Landau Fermi liquid theory for ferromagnets.  Sect.~\ref{s:SpinCurrent} obtains expressions for the spin current in terms of the Fermi liquid theory interaction parameters. Sect.~\ref{s:KE} derives the dynamical equation for the spin current $\vec{J}_{i}$. Sect.~\ref{s:summary} evaluates the two unknown constants in the dynamical equation of $\vec{J}_{i}$ in Ref.~[\onlinecite{SunSaslow24}]. Finally, Sect.~\ref{s:conclusions} presents our conclusions.  An Appendix calculates the magnetic susceptibility of a ferromagnet using Fermi liquid theory.


\section{Density and Polarization Density}
\label{s:density}

This section introduces the basic definitions of density and polarization density in ferromagnets. We employ the notation of Baym and Pethick's authoritative review\cite{BaymPethick78} of Landau Fermi liquid theory (we will refer to this work as BP). 
To avoid possible confusion, we present here some conventions used throughout the paper. Vectors and Greek letters denote spin-space, and Roman indices denote real space.  The symbols $r$ for position and $p$ for momentum represent real-space vectors in the spacial argument of a function, as in a scalar function $g(r,p,t)$. 
When needed, $p_{i}$ denotes the $i$ component of $p$. 

We work with $2\times2$ matrices in spin space for the quasiparticle distribution function $[n_{p}(r,t)]_{\alpha\alpha'}$.  With the Pauli matrices $\vec{\tau}_{\alpha\alpha'}$, and with the variables $r$ and $t$ implicit, following BP (1.1.24) we define
\begin{equation}
\hat{n}_{p}\equiv n_{p\alpha\alpha'}=\tilde{n}_{p}\delta_{\alpha\alpha'}+\vec{\sigma}_{p}\cdot\vec{\tau}_{\alpha\alpha'}.
\label{n_p}
\end{equation}
We use $\tilde{n}_{p}$, not $n_{p}$, to indicate that it is not a variable, but rather a function of the energy, a distinction that is particularly significant later.

From the above definition we have
\begin{eqnarray}
\tilde{n}_{p}(r,t)&=&\frac{1}{2}\sum_{\alpha}[\hat{n}_{p}]_{\alpha\alpha}=\frac{1}{2}{\rm Tr}[\hat{n}],
\label{nscalar}\\
\vec{\sigma}_{p}(r,t)
&=&\frac{1}{2}\sum_{\alpha\alpha'}\vec{\tau}_{\alpha\alpha'}[\hat{n}_{p}]_{\alpha'\alpha}=\frac{1}{2}{\rm Tr}[\vec{\tau} \hat{n}].
\label{nvector}
\end{eqnarray}
This gives
\begin{eqnarray}
\tilde{n}_{p}(r,t)&\equiv&\frac{1}{2}\Big(n_{p\uparrow}(r,t)+n_{p\downarrow}(r,t)\Big)
\label{n_p1},\\
\vec{\sigma}_{p}(r,t)&\equiv&\frac{1}{2}\hat{s}_{p}\Big(n_{p\uparrow}(r,t)-n_{p\downarrow}(r,t)\Big).
\label{sigma_p1}
\end{eqnarray}
where $\hat{s}_{p}(r,t)$ is the local quantization axis for the number $n_{p}$ and polarization $\vec{\sigma}_{p}$. With small deviations preceded by $\delta$, by \eqref{nscalar} we have
\begin{equation}
\delta \tilde{n}_{p}(r,t)\equiv\frac{1}{2}\Big(\delta n_{p\uparrow}(r,t)+\delta n_{p\downarrow}(r,t)\Big).
\label{dn_pF}
\end{equation}
Including a small rotation $\delta\hat{s}_{p}(r,t)$, by \eqref{nvector} we also have
\begin{eqnarray}
\delta \vec{\sigma}_{p}(r,t)&\equiv& \hat{s}_{p}\frac{1}{2}\Big(\delta n_{p\uparrow}(r,t)-\delta n_{p\downarrow}(r,t)\Big)\cr
&&+\delta\hat{s}_{p}\frac{1}{2}\Big(n_{p\uparrow}(r,t)- n_{p\downarrow}(r,t)\Big) .
\label{dsigma_pF}
\end{eqnarray}
%
The total number density then is
\begin{equation}
\tilde{n}(r,t)= \frac{2}{V}\sum_{p}\tilde{n}_{p}(r,t)=\int d\tau \Big( n_{p\uparrow}(r,t)+n_{p\downarrow}(r,t) \Big),
\label{n}
\end{equation}
and the total polarization density is
\begin{equation}
\vec{\sigma}(r,t)= \frac{2}{V}\sum_{p}\vec{\sigma}_{p}(r,t)=\int d\tau \hat{s}_{p}\Big( n_{p\uparrow}(r,t)-n_{p\downarrow}(r,t) \Big),
\label{sigma}
\end{equation}
where we replace the summations over $p$ as integrations:
\begin{equation}
\frac{1}{V}\sum_{p}\rightarrow \int\frac{d^{3}p}{(2\pi\hbar)^{3}} \equiv \int d\tau .
\label{tau}
\end{equation}
With $\gamma$ the gyromagnetic ratio, and with spin in units of $\hbar/2$, the magnetization (magnetic moment density) is given by
\begin{equation}
\vec{M}=\frac{\gamma\hbar}{2}\vec{\sigma}.
\label{M}
\end{equation}

\section{Landau Fermi Liquid Theory}
\label{s:LFLT}


Landau Fermi liquid theory has two parts. In the first part the excitations are given interaction energies that reflect the presence of other excitations.  In the second part the kinetic theory of the excitation number is considered.

Landau uses $E$ for the energy density, as do BP.  For the individual excitation energy matrix we follow BP (1.1.23) (but add a tilde where appropriate).  Thus we define
\begin{equation}
\hat{\epsilon}_{p}\equiv \epsilon_{p \alpha\alpha'}=\tilde{\epsilon}_{p}\delta_{\alpha\alpha'}+\vec{h}_{p}\cdot\vec{\tau}_{\alpha\alpha'}.
\label{eps_pmat}
\end{equation}
Typically the direction of $\vec{h}_{p}$, given by $\hat{s}'_{p}$, differs from the direction of $\vec{\sigma}_{p}$, given by $\hat{s}_{p}$, so we write
\begin{eqnarray}
\tilde{\epsilon}_{p}&=&\frac{1}{2}(\epsilon_{p\uparrow}+\epsilon_{p\downarrow}),
\label{eps_p}\\
\vec{h}_{p}&=&\frac{1}{2}\hat{s}'_{p}(\epsilon_{p\uparrow}-\epsilon_{p\downarrow}).
\label{vech_p}
\end{eqnarray}
In practice this distinction will not be shown explicitly.

Both $\epsilon_{p}(r,t)$ and $\vec{h}_{p}(r,t)$ have parts that involve no deviations from local equilibrium, with subscript 0, and deviations from local equilibrium, with prefix $\delta$. With
\begin{equation}
E=E_{0}+\delta E+\frac{1}{2}\delta^{2}E,
\label{E}
\end{equation}
BP (1.1.22) and (1.1.24) give
\begin{eqnarray}
\delta E&=&\frac{1}{V}\sum_{p}\sum_{\alpha\alpha'}(\epsilon_{p})_{\alpha\alpha'}(\delta\tilde{n}_{p})_{\alpha'\alpha},
\label{delE}\\
\delta^{2}E&=&\frac{1}{V^{2}}\sum_{pp'}\sum_{\alpha\sigma\alpha'\sigma'}f_{p\alpha\sigma,p'\alpha'\sigma'}(\delta n_{p\alpha\sigma})(\delta n_{p'\alpha'\sigma'}), \quad
\label{del2E}
\end{eqnarray}
where $f_{p\alpha\sigma,p'\alpha'\sigma'}$ describes the Fermi liquid interaction between excitations. BP (1.1.25) write $f$, with units of (energy)$\times$(volume) as
\begin{equation}
f_{p\alpha\sigma,p'\alpha'\sigma'}=f^{(s)}_{pp'}\delta_{\alpha\sigma}\delta_{\alpha'\sigma'}+f^{(a)}_{pp'}\vec{\tau}_{\alpha\sigma}\cdot\vec{\tau}_{\alpha'\sigma'},
\label{f2}
\end{equation}
where the superscripts $(s)$ and $(a)$ refer to symmetric and antisymmetric. In implicit $(2,2)\times(2,2)$ bi-matrix notation this is
\begin{equation}
\hat{f}_{pp'}=f^{(s)}_{pp'}+f^{(a)}_{pp'}\vec{\tau}\cdot\vec{\tau'}.
\label{f3}
\end{equation}

For the interaction energy $\delta\epsilon_{p\alpha\beta}$ we have
\begin{eqnarray}
\delta\epsilon_{p\alpha\beta}&\equiv& \frac{V}{2}\frac{\partial}{\partial n_{p\alpha\beta}}(\delta^{2}E) \cr
&=&\frac{1}{V}\sum_{p'\alpha'\beta'}f_{p\alpha\beta,p'\alpha'\beta'}\delta n_{p'\alpha'\beta'}.
\label{deleps1}
\end{eqnarray}
This is consistent, in spin-diagonal form, with the second term in BP (1.1.17). Note the Pauli matrix property
\begin{equation}
\tau_{\alpha\beta',j} \tau_{\beta'\alpha',i}=\delta_{ij}\delta_{\alpha\alpha'}+i\epsilon_{jik}\tau_{\alpha\alpha',k}.
\label{taus}
\end{equation}
Setting $\alpha=\alpha'$ in \eqref{taus} and summing over $\alpha$ then gives
$\tau_{\alpha'\beta',j} \tau_{\beta'\alpha',i}=2\delta_{ij}$. We now rewrite \eqref{deleps1} using \eqref{f2}, \eqref{taus} and the variation of \eqref{n_p}.  This gives
\begin{equation}
\delta\epsilon_{p\alpha\beta}=\frac{2}{V}\sum_{p'}\Big(f^{(s)}_{pp'}\delta \tilde{n}_{p'}\delta_{\alpha\beta}+ f^{(a)}_{pp'}\delta \vec{\sigma}_{p'}\cdot\vec{\tau}_{\alpha\beta}\Big).
\label{deleps2}
\end{equation}

Following BP (1.1.27) and (1.1.28) we write
\begin{eqnarray}
\tilde{\epsilon}_{p}&=&\tilde{\epsilon}^{(0)}_{p}(r,t)+\frac{2}{V}\sum_{p'}f^{(s)}_{pp'}\delta\tilde{n}_{p'}\cr
&\equiv&\tilde{\epsilon}^{(0)}_{p}(r,t)+\delta\tilde{\epsilon}_{p}(r,t), \qquad
\label{eps_p1} \\
\vec{h}_{p}&=&\vec{h}^{(0)}_{p}+\frac{2}{V}\sum_{p'}f^{(a)}_{pp'}\delta \vec{\sigma}_{p'}\cr
&\equiv&\vec{h}^{(0)}_{p}+\delta\vec{h}_{p}.
\label{vech_p1}
\end{eqnarray}
For a paramagnet with negative gyromagnetic ratio and taking $\gamma>0$, we have $\vec{h}_{p}^{(0)}=-(\gamma\hbar/2)\vec{B}$.  For a ferromagnet in its ground state, with magnetization direction $\hat{M}$,
\begin{eqnarray}
\vec{h}_{p}^{(0)}&=&- \frac{1}{2}\gamma\hbar  \vec{B}+\frac{2}{V}f^{(a)}_{0}\sum_{p}\vec{\sigma}_{p} \cr
&=&- \frac{1}{2}\gamma\hbar  \vec{B}+f^{(a)}_{0}\vec{\sigma}. 
\label{h^0a}
\end{eqnarray}
This assumes the same interactions for the excitations as for the ground state. Comparison of  \eqref{deleps2} with \eqref{eps_p1} and \eqref{vech_p1} gives the quasiparticle interaction terms
\begin{eqnarray}
\delta\tilde{\epsilon}_{p} &=&\frac{2}{V}\sum_{p'}f^{(s)}_{pp'}\delta \tilde{n}_{p'}=2\int d\tau f^{(s)}_{pp'}\delta\tilde{n}_{p'},
\label{eps_p2}\\
\delta\vec{h}_{p}&=&\frac{2}{V}\sum_{p'}f^{(a)}_{pp'}\delta \vec{\sigma}_{p'}=2 \int d\tau' f^{(a)}_{pp'}\delta \vec{\sigma}_{p'}.
\label{vech_p2}
\end{eqnarray}

\section{Current and Spin Current}
\label{s:SpinCurrent}
This section derives expressions for the spin current density in terms of the Fermi liquid theory parameters. For completeness, we first discuss the (number) current density.

\subsection{Number Current Density}
\label{ss:Current}
The number current density along $i$ for a quasiparticle with momentum $p$ is
\begin{eqnarray}
J_{pi}(r,t)&=&\frac{1}{V}\sum_{\alpha\alpha'} [n_{p}(r,t)]_{\alpha\alpha'} \frac{\partial[\epsilon_{p}(r,t)]_{\alpha'\alpha}}{\partial p_{i}} \cr
&=&\frac{2}{V}\Big(\frac{\partial\tilde{\epsilon}_{p}}{\partial p_{i}}\tilde{n}_{p}+ \frac{\partial \vec{h}_{p}}{\partial p_{i}}\cdot\vec{\sigma_{p}}\Big).
\label{j_pi}
\end{eqnarray}
Observe that $\vec{\sigma}_{p}$ and $\partial_{p_i}\vec{h}_{p}$ can have different directions.  For ferromagnets with common axis $\hat{s}$, small deviations from $\hat{s}$ give $\vec{\sigma}_{p}$ and $\partial_{p_i}\vec{h}_{p}$ a deviation from alignment that is second order in the misalignment angle, so for longitudinal components the deviation from alignment may often be neglected.

Summing over $p$ we then have
\begin{equation}
J_{i}(r,t)={\hskip-0.1cm} \int {\hskip-0.1cm} d\tau J_{pi}= 2  \int    d\tau \Big(\frac{\partial\tilde{\epsilon}_{p}}{\partial p_{i}}\tilde{n}_{p}+\frac{\partial \vec{h}_{p}}{\partial p_{i}}\cdot\vec{\sigma_{p}}\Big).
\label{j_i0}
\end{equation}
With the diagonal spin direction $\hat{s}_{p}$ for $p$, we have
\begin{equation}
n_{p\uparrow}=n_{p}+\vec{\sigma}_{p}\cdot\hat{s}_{p}, \quad n_{p\downarrow}=n_{p}-\vec{\sigma}_{p}\cdot\hat{s}_{p}.
\label{nupdown_pi]}
\end{equation}
Defining the velocities\cite{BaymPethick78}
\begin{eqnarray}
 v_{ p \uparrow_{i}}=\frac{\partial\epsilon_{p\uparrow}}{\partial p_{i}}
&=&\frac{\partial\tilde{\epsilon}_{p}}{\partial p_{i}}+ \Big(\frac{\partial\vec{h}_{p}}{\partial p_{i}}\cdot\hat{s}_{p}\Big), \\
 v_{ p\downarrow_{i}}=\frac{\partial\epsilon_{p\downarrow}}{\partial p_{i}}
&=&\frac{\partial\tilde{\epsilon}_{p}}{\partial p_{i}}- \Big(\frac{\partial\vec{h}_{p}}{\partial p_{i}}\cdot\hat{s}_{p}\Big),
\label{vupdown_pi}
\end{eqnarray}
it is perhaps more transparent to write
\begin{eqnarray}
J_{pi}(r,t)&=&\frac{\partial\epsilon_{p\uparrow}}{\partial p_{i}} n_{p\uparrow}+\frac{\partial\epsilon_{p\downarrow}}{\partial p_{i}} n_{p\downarrow}, \cr
&=&v_{pi\uparrow}n_{p\uparrow}+v_{pi\downarrow}n_{p\downarrow},
\label{j_piF}
\end{eqnarray}
where the Fermi surface information is contained in the distribution function. The total current density can then be written as
\begin{equation}
J_{i}(r,t)=\int d\tau \Big(\frac{\partial\epsilon_{p\uparrow}}{\partial p_{i}} n_{p\uparrow}+\frac{\partial\epsilon_{p\downarrow}}{\partial p_{i}} n_{p\downarrow}\Big).
\label{j_i}
\end{equation}


\subsection{Spin Current Density}
\label{ss:SpinCurrent}
The spin current density along $i$ (recall that vector arrows imply spin space) for a quasiparticle $p$ is
\begin{eqnarray}
\vec{J}_{pi}(r,t)&=&\frac{1}{V}\sum_{\alpha\alpha'\alpha''} [n_{p}(r,t)]_{\alpha\alpha'} \frac{\partial [\epsilon_{p}(r,t)]_{\alpha'\alpha''}}{\partial p_{i}}\vec{\tau}_{\alpha''\alpha} \quad\\
&=&\frac{2}{V}\Big(\frac{\partial\tilde{\epsilon}_{p}}{\partial p_{i}}\vec{\sigma}_{p}(r,t)+ \frac{\partial \vec{h}_{p}}{\partial p_{i}}\tilde{n}_{p}(r,t)\Big).
\label{vecJ_pi1}
\end{eqnarray}

Following BP (1.3.77),
the total spin current density along $i$ is
\begin{eqnarray}
\vec{J}_{i}(r,t)&=&2\int d\tau \Big[\frac{\partial\tilde{\epsilon}_{p}}{\partial p_{i}}\vec{\sigma}_{p}(r,t)+ \frac{\partial \vec{h}_{p}}{\partial p_{i}}\tilde{n}_{p}(r,t) \Big] \cr
&\equiv& 2\int d\tau \vec{J}_{pi}(r,t). \qquad
\label{vecJ_pi2}
\end{eqnarray}
In the above the $\partial/\partial p_{i}$ terms have spin indices determined by the respective factors $\vec{\sigma}_{p}$ and $\tilde{n}_{p}$.  

For the paramagnet, BP take $\vec{\sigma}_{p}$ to be small, even in equilibrium.  For a ferromagnet $\vec{\sigma}_{p}$ has an equilibrium component $\vec{\sigma}^{(0)}_{p}$, leading to a ground state exchange field.  
Thus the two terms in \eqref{vecJ_pi2} 
may be approximated by
\begin{eqnarray}
\frac{\partial\tilde{\epsilon}_{p}}{\partial p_{i}}\vec{\sigma}_{p}(r,t)&\approx&
\frac{\partial\delta\tilde{\epsilon}_{p}}{\partial p_{i}}\vec{\sigma}^{(0)}_{p}(r,t)+\frac{\partial\tilde{\epsilon}^{(0)}_{p}}{\partial p_{i}}\delta\vec{\sigma}_{p}(r,t);
\label{A}\\
\frac{\partial \vec{h}_{p}}{\partial p_{i}}\tilde{n}_{p}(r,t)&\approx&
\frac{\partial \delta\vec{h}_{p}}{\partial p_{i}}\tilde{n}^{(0)}_{p}(r,t)+\frac{\partial \vec{h}^{(0)}_{p}}{\partial p_{i}}\delta \tilde{n}_{p}(r,t).
\label{B}
\end{eqnarray}
In \eqref{A} the first of the terms is not present for a paramagnet, and in \eqref{B} the second of the terms also is not present for a paramagnet.  
These two new terms both give a component to $\vec{J}_{i}(r,t)$ that is along $\vec{M}$, whereas our interest is in the transverse $\vec{J}_{i}(r,t)$.

As done for the paramagnet by Leggett and by BP, we can combine the second term of \eqref{A} with an integration by parts on the first term of \eqref{B}.  On doing the sum and suppressing $(r,t)$ on the right hand side, for the transverse spin components we obtain:
\begin{eqnarray}
\vec{J}_{i}(r,t)&=&2\int d\tau \left(\frac{\partial \tilde{\epsilon}^{(0)}_{p}}{\partial p_{i}}\delta\vec{\sigma}_{p}
-\delta\vec{h}_{p}\frac{\partial \tilde{n}^{(0)}_{p}}{\partial p_{i} } \right).
\label{vecJ_pi3}
\end{eqnarray}

Below we evaluate the two terms in \eqref{vecJ_pi3}. For the first term, we have
\begin{equation}
\vec{J}^{(1)}_{i}(r,t)=2 \int d\tau \frac{\partial \tilde{\epsilon}^{(0)}_{p} }{ \partial p_{i} }\delta\vec{\sigma}_{p}
=2 \int d\tau \frac{\partial \tilde{\epsilon}^{(0)}_{p} }{ \partial p }\hat{p}_{i}\delta\vec{\sigma}_{p}.
\label{Jvec1}
\end{equation}
The second term of \eqref{vecJ_pi3} contains
$\tilde{n}_{p}$, a function of the variable $\epsilon_{p}$.  We should not confuse the integration variable $\epsilon_{p}$ with the function $\tilde{\epsilon}_{p}=(1/2)(\epsilon_{p\uparrow}+\epsilon_{p\downarrow})$.  Then, with $\tilde{n}_{p}=(1/2)(n_{p\uparrow}+n_{p\downarrow})$, we have
\begin{equation}
\frac{\partial \tilde{n}_{p}}{\partial p}=\frac{\partial \tilde{n}_{p}}{\partial \epsilon_{p}}\frac{\partial \epsilon_{p}}{\partial p}
=\frac{1}{2}\left(\frac{\partial n_{p\uparrow}}{\partial \epsilon_{p\uparrow}}\frac{\partial \epsilon_{p\uparrow}}{\partial p} + \frac{\partial n_{p\downarrow}}{\partial \epsilon_{p\downarrow}}\frac{\partial \epsilon_{p\downarrow}}{\partial p}\right).
\label{tilden-eps}
\end{equation}
Substituting \eqref{vech_p2} into the second term of \eqref{vecJ_pi3} gives
\begin{eqnarray}
\vec{J}^{(2)}_{i}(r,t)
&=&-4\int d\tau \frac{\partial \tilde{n}^{(0)}_{p} }{ \partial \epsilon_{p} }\frac{ \partial \epsilon_{p} }{ \partial p }\hat{p}_{i}
	\int d\tau' f^{(a)}_{pp'}\delta\vec{\sigma}_{p'}.
\label{Jvec2}
\end{eqnarray}

We now employ $f_{p'p}\approx f^{(a)}_{0}+f^{(a)}_{1}\hat{p}'_{j}\hat{p}_{j}$.  The angular part of $\int \hat{p}_{i}d\tau$ eliminates $f^{(a)}_{0}$ and the angular part of $\int \hat{p}_{i}\hat{p}_{j}d\tau$ replaces $\hat{p}_{i}\hat{p}'_{j}\hat{p}_{j}$ by $(1/3)\hat{p}'_{i}$, to give a term $(1/3)f^{(a)}_{1}\hat{p}'_{i}$. The $\int d\tau$ term on $-(\partial \tilde{n}^{(0)}_{p}/ \partial \epsilon_{p})( \partial \epsilon_{p} / \partial p )$ then gives $(1/2)(N_{\uparrow}v_{F\uparrow}+N_{\downarrow}v_{F\downarrow})$ according to \eqref{tilden-eps}, where the $N_{\uparrow\downarrow}$'s are the up and down spin densities of states, defined explicitly in \eqref{Ntildes}.  Thus
\begin{equation}
\vec{J}^{(2)}_{i}(r,t)= \frac{2}{3}f^{(a)}_{1}(N_{\uparrow}v_{F\uparrow}+N_{\downarrow}v_{F\downarrow}) \int d\tau \hat{p}_{i}\delta\vec{\sigma}_{p}.
\label{Jvec2b}
\end{equation}
With total density of states $N(0)=N_{\uparrow}+N_{\downarrow}$, in the paramagnetic limit $v_{F\uparrow}, v_{F\downarrow}\rightarrow v_{F}$, 
so $N_{\uparrow}v_{F\uparrow}+N_{\downarrow}v_{F\downarrow}\rightarrow v_{F}N(0)$. 
This matches the second term of BP (1.3.79) where $F^{(a)}_{l}\equiv f^{(a)}_{l}N(0)$.

Thus, summing \eqref{Jvec1} and \eqref{Jvec2b} gives
\begin{equation}
\vec{J}_{i}(r,t)=2u\int d\tau \hat{p}_{i}\delta\vec{\sigma}_{p},
\label{Jvec4}
\end{equation}
where the constant $u$, with units of velocity, is introduced to simplify the equations:
\begin{equation}
u=\frac{1}{2}\Big[(v_{F\uparrow}(1+\frac{2}{3}f^{(a)}_{1}N_{\uparrow})+v_{F\downarrow}(1+\frac{2}{3}f^{(a)}_{1}N_{\downarrow})\Big].
\label{C}
\end{equation}
In the paramagnetic limit, where $N_{\uparrow}=N_{\downarrow}=N(0)/2$, $u$ goes to the paramagnetic value
 \begin{align}
u^{P}&=v_{F}\left(1+\frac{1}{3}f^{(a)}_{1}N(0)\right)=v_{F}\left(1+\frac{1}{3}F^{(a)}_{1}\right)\nn\\
&=v_{F}\left(1+\frac{1}{12}Z_{1}\right);
\label{CP}
\end{align}
following BP we employ $F^{(a)}_{l}\equiv Z_{l}/4$.  The result \eqref{CP} is implicit in BP (1.3.79) and in Leggett (20). For comparison to the results in Leggett,\cite{Leggett70} note that in Leggett's notation $\zeta_{l}=(dn/d\epsilon)^{-1}Z_{l}$.  Then, since $N(0)=dn/d\epsilon$ and $F^{(a)}_{l}\equiv f^{(a)}_{l}N(0)$, we have $4f^{(a)}_{l}=4F^{(a)}_{l}/N(0)=Z_{l}/N(0)=\zeta_{l}$.


\section{Kinetic Equation}
\label{s:KE}

This section derives the equation of motion $\partial\vec{J}_{i}/\partial t$ for the spin current density \eqref{Jvec4}. We employ Silin's kinetic equation for $\partial_{t}\vec{\sigma}_{p}$, i.e. Eq. (1.8) in [\onlinecite{Silin1}]. Rewriting it in our notation with $f_{p}\rightarrow \tilde{n}_{p}$, $\epsilon_{1}\rightarrow\tilde{\epsilon}_{p}$, $\vec{\sigma}\rightarrow\vec{\sigma}_{p}$, and $\vec{\epsilon}_{2}\rightarrow\vec{h}_{p}$, we have
\begin{eqnarray}
&&\frac{\partial \vec{\sigma}_{p}}{\partial t}
+
\frac{\partial \tilde{\epsilon}_{p}}{\partial p_{i}}\frac{\partial \vec{\sigma}_{p}}{\partial r_{i}}
-\frac{\partial \tilde{\epsilon}_{p}}{\partial r_{i}}\frac{\partial \vec{\sigma}_{p}}{\partial  p_{i}}
+\frac{\partial \tilde{n}_{p}}{\partial  r_{i}}\frac{\partial \vec{h}_{p}}{\partial p_{i}}
-\frac{\partial \tilde{n}_{p}}{\partial p_{i}}\frac{\partial \vec{h}_{p}}{\partial r_{i}}\cr
&&=-\frac{2}{\hbar}\vec{\sigma}_{p}\times\vec{h}_{p} +\frac{\partial \vec{\sigma}_{p}}{\partial t}\Big |_{c}.\qquad
\label{kineq-sigp2}
\end{eqnarray}
The sign of the $\vec{\sigma}_{p}\times\vec{h}_{p}$ term differs from that in Silin (1.8),\cite{Silin1} but agrees with that in BP (1.3.60).

The last term is the collision term, which must be put in by hand, but is not of present interest. Here $\delta\tilde{\epsilon}_{p}$, $\partial r_{i}$, $\vec{\sigma}_{p}$, and $\vec{h}_{p}$ are of first order.  Following BP, we now include only terms of second order or less in deviations from equilibrium.
As for the paramagnet (in BP), for the ferromagnet the third term on the left is third order ($\partial \epsilon_{p}/\partial r_{i}$ is second order and $\vec{\sigma}_{p}$ is first order), and the fourth term is third order ($\partial \vec{h}_{p}/\partial p_{i}$ is first order, and $\partial\vec{\sigma}_{p}/\partial r_{i}$ is second order).

With $\partial \tilde{n}_{p}/\partial p_{i}=(\partial \tilde{n}_{p}/\partial \epsilon_{p})(\partial \epsilon_{p}/\partial p_{i})$, for the transverse part we obtain the same equation as BP for the paramagnet, or
\begin{equation}
\frac{\partial \vec{\sigma}_{p}}{\partial t}
+\frac{\partial \tilde{\epsilon}^{(0)}_{p}}{\partial p_{i}}\frac{\partial \vec{\sigma}_{p}}{\partial r_{i}}
-\frac{\partial \tilde{n}^{(0)}_{p}}{\partial p_{i}}\frac{\partial \vec{h}_{p}}{\partial r_{i}}
\approx-\frac{2}{\hbar}\vec{\sigma}_{p}\times\vec{h}_{p} +\frac{\partial \vec{\sigma}_{p}}{\partial t}\Big |_{c}.\quad
\label{kineq-sigp3}
\end{equation}
Now let's define
\begin{equation}
\tilde{v}^{(0)}_{pi}=\frac{\partial \tilde{\epsilon}^{(0)}_{p}}{\partial p_{i}}=\frac{1}{2}\frac{\partial (\epsilon^{(0)}_{p\uparrow}+\epsilon^{(0)}_{p\downarrow})}{\partial p_{i}},
\label{tildev}
\end{equation}
and
\begin{eqnarray}
\frac{\partial \tilde{n}^{(0)}_{p}}{\partial p_{i}}&=&\frac{1}{2}\frac{\partial(n^{(0)}_{p\uparrow}+n^{(0)}_{p\downarrow})}{\partial p_{i}}
=\frac{1}{2}\left(\frac{\partial n^{(0)}_{p\uparrow}}{\partial \epsilon_{p\uparrow}}\frac{\partial\epsilon^{(0)}_{p\uparrow}}{\partial p_{i}}
+\frac{\partial n^{(0)}_{p\downarrow}}{\partial\epsilon_{p\downarrow}}\frac{\partial\epsilon^{(0)}_{p\downarrow}}{\partial p_{i}}\right)\cr
&=&\Bigl[ \frac{\partial \tilde{n}^{(0)}_{p}}{\partial \epsilon^{(0)}_{p}}\frac{\partial \tilde{\epsilon}^{(0)}_{p}}{\partial p_{i}} \Bigr]_{\rm dd}
=\Bigl[ \tilde{v}^{(0)}_{pi} \frac{\partial \tilde{n}^{(0)}_{p}}{\partial \epsilon_{p} } \Bigr]_{\rm dd},
\label{redodndpi}
\end{eqnarray}
where $[]_{\rm dd}$ means ``double diagonal'', i.e. taking only take the two diagonal terms. Without the $[]_{\rm dd}$ notation the above would be the average of four, not two, terms.  

We now rewrite \eqref{kineq-sigp3} as
\begin{equation}
\frac{\partial \vec{\sigma}_{p}}{\partial t}
+\Bigl[ \tilde{v}^{(0)}_{pi}\frac{\partial }{\partial r_{i}}(\vec{\sigma}_{p}-\frac{\partial \tilde{n}^{(0)}_{p}}{\partial \epsilon_{p}} \vec{h}_{p}) \Bigr]_{\rm dd}
\approx-\frac{2}{\hbar}\vec{\sigma}_{p}\times\vec{h}_{p} +\frac{\partial \vec{\sigma}_{p}}{\partial t}\Big |_{c}.\qquad\qquad
\label{kineq-sigp4}
\end{equation}
This is similar to BP (1.3.80), where the double diagonal restriction is not needed.  Applying \eqref{Jvec4} to do $\int d\tau$ on \eqref{kineq-sigp4}, we arrive at a  dynamical equation for the spin current that has the same form as BP (1.3.81): 
\begin{equation}
\frac{\partial }{\partial t}\vec{J}_{i}+\frac{\partial }{\partial r_{k}}\vec{\Pi}_{ik}=\frac{\partial }{\partial t}\vec{J}_{i}|_{p}+\frac{\partial }{\partial t}\vec{J}_{i}|_{c},
\label{dJ/dt}
\end{equation}
with subscripts $p$ (precession) and $c$ (collision).  In \eqref{dJ/dt} we call the term in $\partial_{k}\vec{\Pi}_{ik}$ the magnetic (or spin) pressure term and the 
term in $ \partial_t\vec{J}_{i}|_{p}$ the precession term, respectively.

The second term on the left-hand-side of \eqref{kineq-sigp4} gives the magnetic pressure term, where by \eqref{Jvec4} we include a factor of $2C\int d\tau \hat{p}_{i}$ to obtain
\begin{equation}
\vec{\Pi}_{ik}= 2u \int d\tau \hat{p}_{i} \Big[ \tilde{v}^{(0)}_{pk} (\vec{\sigma}_{p}-\frac{\partial \tilde{n}^{(0)}_{p}}{\partial \epsilon_{p}} \vec{h}_{p}) \Big]_{\rm dd}.
\label{Pi-ik}
\end{equation}
As with BP for the paramagnet, we expect that the transverse component satisfies $\vec{\Pi}_{ik,\perp}\sim \delta_{ik}(\delta\vec{\sigma})_{\perp}\sim \delta_{ik}\delta\vec{M}_{\perp}$.

The third term in \eqref{kineq-sigp4} gives,  on doing $\int d\tau$, the precession term.  By \eqref{Jvec4} we include a factor of $2u\int d\tau \hat{p}_{i}$ to obtain
\begin{equation}
\frac{\partial }{\partial t}\vec{J}_{i}|_{p}=-\frac{4u}{\hbar}\int d\tau \hat{p}_{i}\vec{\sigma}_{p}\times\vec{h}_{p}.
\label{vecJiprecess}
\end{equation}
Below we evaluate the magnetic pressure and the precession terms.

\subsection{Magnetic Pressure Term}
\label{s:MagPressure}
The $\vec{\sigma}_{p}$ term in \eqref{Pi-ik} gives, after an angular average, and $\vec{\sigma}^{(0)}=\sigma^{(0)}\hat{M}=2\int d\tau\vec{\sigma}_{p}$,
\begin{eqnarray}
\vec{\Pi}^{(1)}_{ik} =  2u \int d\tau \hat{p}_{i} \tilde{v}^{(0)}_{pk} \vec{\sigma}_{p}
= \frac{1}{6}u (v_{F\uparrow}+v_{F\downarrow}) \delta_{ik} \vec{\sigma}^{(0)}.
\label{Pi-ik1}
\end{eqnarray}
By \eqref{vech_p1} we have $\vec{h}_{p}=2\int d\tau' f^{(a)}_{pp'}\vec{\sigma}_{p'}$, so the $\vec{h}_{p}$ term in \eqref{Pi-ik} gives
\begin{eqnarray}
\vec{\Pi}^{(2)}_{ik}&=& -2u \int d\tau \hat{p}_{i} \big[ \tilde{v}^{(0)}_{pk} \frac{\partial \tilde{n}^{(0)}_{p}}{\partial \epsilon_{p}} \vec{h}_{p}\big]_{\rm dd},\cr
&=&-\frac{4}{3}u\delta_{ik}\int d\tau \big[ \tilde{v}^{(0)}_{p} \frac{\partial \tilde{n}^{(0)}_{p}}{\partial \epsilon_{p}} \big]_{\rm dd} \int d\tau' f^{(a)}_{pp'}\vec{\sigma}_{p'}.
\end{eqnarray}
Expanding the terms in $[]_{\rm dd}$ gives
\begin{eqnarray}
\vec{\Pi}^{(2)}_{ik}&=&-\frac{2}{3}u \delta_{ik} {\hskip-0.1cm} \int  {\hskip-0.1cm} d\tau \left( v^{(0)}_{p\uparrow}\frac{\partial \tilde{n}^{(0)}_{p\uparrow}}{\partial \epsilon_{p\uparrow}}
+v^{(0)}_{p\downarrow}\frac{\partial \tilde{n}^{(0)}_{p\downarrow}}{\partial \epsilon_{p\downarrow}}\right)  {\hskip-0.1cm} \int d\tau' {\hskip-0.1cm} f^{(a)}_{pp'}\vec{\sigma}_{p'}\cr
&=&\frac{1}{3}u f^{(a)}_{0}\delta_{ik} \left(v_{F\uparrow}N_{\uparrow}+v_{F\downarrow}N_{\downarrow} \right) \vec{\sigma}^{(0)}.
\label{Pi-ik2}
\end{eqnarray}
Thus 
the sum of \eqref{Pi-ik1} and \eqref{Pi-ik2} yields
\begin{eqnarray}
\vec{\Pi}_{ik}{\hskip-0,18cm}&=&{\hskip-0,18cm} \frac{1}{6}u \delta_{ik} \Big[(v_{F\uparrow}{\hskip-0,05cm}+v_{F\downarrow}){\hskip-0,05cm}+{\hskip-0,05cm}2f^{(a)}_{0}(v_{F\uparrow}N_{\uparrow}{\hskip-0,05cm}+{\hskip-0,05cm}v_{F\downarrow}N_{\downarrow}) \Big] \vec{\sigma}^{(0)}.\qquad
\label{Pi-ik3}
\end{eqnarray}
In the paramagnetic limit this goes to
\begin{eqnarray}
\vec{\Pi}_{ik}^{P}
&=& \frac{1}{3}u^{P}v_{F} \delta_{ik} \left(1+ f^{(a)}_{0}N(0) \right) \vec{\sigma}^{(0)}.
\label{Pi-ik3P}
\end{eqnarray}
Note that $f^{(a)}_{0}N(0)$ in \eqref{Pi-ik3P} can be rewritten in various notations: $f^{(a)}_{0}N(0)= F^{(a)}_{0}=  Z_{0}/4$. Recalling the definition of $u_P$ in \eqref{CP}, this agrees with BP (1.3.87) and with Leggett (20).


\subsection{Precession Term}
\label{s:precess}
BP (1.3.89) gives the paramagnet precession term, to which we will compare after taking the paramagnetic limit of the results for the ferromagnet.  BP argue that in the cross-product of \eqref{vecJiprecess}, on writing $\vec{h}_{p}=2\int d\tau' f^{(a)}_{pp'}\vec{\sigma}_{p'}$, the cross-product involves only the $l=0$ part of $\vec{\sigma}_{p}$ and the $l=1$ part of $\vec{\sigma}_{p'}$, and vice-versa.  We can write the right hand side of \eqref{vecJiprecess} as $-(4u/\hbar)\vec{I}_{i}$, where
\begin{eqnarray}
\vec{I}_{i}&=&2\int d\tau \int d\tau' \hat{p}_{i} (f^{(a)}_{0}+f^{(a)}_{1}\hat{p}_{j}\hat{p}'_{j}+\dots)\vec{\sigma}_{p}\times\vec{\sigma}_{p'} \cr
&=&\frac{1}{u}\left(f^{(a)}_{0}-\frac{1}{3}f^{(a)}_{1}\right)\vec{J}_{i}\times \vec{\sigma}^{(0)}.
\label{auxpre}
\end{eqnarray}
In the above we used $\vec{J}_{i}=2u\int d\tau \hat{p}_{i}\vec{\sigma}_{p}$.  The higher order $f^{(a)}_{l}$ do not contribute because there are only two ways for the $f^{(a)}_{l}$ to appear in $\vec{I}_{i}$. Then
\begin{equation}
\frac{\partial }{\partial t}\vec{J}_{i}|_{p}=-\frac{4u}{\hbar}\vec{I}_{i}=-\frac{4}{\hbar}\left(f^{(a)}_{0}-\frac{1}{3}f^{(a)}_{1}\right)\vec{J}_{i}\times \vec{\sigma}^{(0)}.
\label{precessf}
\end{equation}
%
%
This agrees with Leggett (20).\cite{Pethickprivate}  

\section{Determination of $G$ and $\lambda$
}
\label{s:summary}
We now evaluate the unknown constants for vector spin pressure term and the mean-field precession term in Ref.~[\onlinecite{SunSaslow24}], namely, $G$ and $\lambda$ in \eqref{dJM/dt-time}.
Note that $G$ has units of velocity squared and $\lambda$ has units of $\mu_{0}$. From \eqref{Pi-ik3}, we obtain
\begin{eqnarray}
G{\hskip-0.4cm}&=&{\hskip-0.4cm} 
 \frac{1}{6} u\Big[(v_{F\uparrow}+v_{F\downarrow})+2f^{(a)}_{0}(v_{F\uparrow}N_{\uparrow}+v_{F\downarrow}N_{\downarrow}) \Big]\nn\\
&=&{\hskip-0.4cm}\frac{1}{12} \Big[(v_{F\uparrow}(1+\frac{2}{3}f^{(a)}_{1}N_{\uparrow})+v_{F\downarrow}(1+\frac{2}{3}f^{(a)}_{1}N_{\downarrow})\Big]\nn\\
&\qquad\times&{\hskip-0.2cm} \Big[(v_{F\uparrow}+v_{F\downarrow})+2f^{(a)}_{0}(v_{F\uparrow}N_{\uparrow}+v_{F\downarrow}N_{\downarrow}) \Big],
\label{GFL}
\end{eqnarray}
where $u$ is given in \eqref{C}, and recall that $\vec{M}=(\gamma\hbar/2)\vec{\sigma}$. In the paramagnetic limit, this gives
\begin{eqnarray}
G^P&=& 
 \frac{1}{3} u^P v_F \left(1+F^{(a)}_{0}\right)\nn\\
&=&  \frac{1}{3}v_{F}^2\left(1+\frac{1}{3}F^{(a)}_{1}\right)  \left(1+F^{(a)}_{0}\right),
\label{}
\end{eqnarray}
where $u^P$ is given in \eqref{CP}.
This is a very complex form, containing not only the ferromagnetic parameters $v_{\uparrow,\downarrow}$ and $N_{\uparrow,\downarrow}$, but also the ferromagnetic Fermi liquid parameters $f^{(a)}_{0,1}$.

From \eqref{precessf}, we obtain
\begin{eqnarray}
\lambda&=&\frac{8}{\gamma^2 \hbar^2} \left(f^{(a)}_{0}-\frac{1}{3}f^{(a)}_{1}\right).
\label{lambdaFL}
\end{eqnarray}
This has a relatively simple form, depending only on the $f^{(a)}_{0,1}$.  Eq. \eqref{chipar} of the Appendix shows that the magnetic susceptibility depends on $N_{\uparrow}$, $N_{\downarrow}$, and $f_{0}^{(a)}$.

Since  $f$ has units [(energy)$\times$(volume)], $\lambda$ in \eqref{lambdaFL} has units [(energy)$\times$(volume)$/(\gamma \hbar)^2$] or [(energy)$/$((volume)$M^2$)]. This is the same as [$\mu_{0}$], since $\mu_0 M^2$ has units [(energy)$/$(volume)].


\section{Conclusions}
\label{s:conclusions}
This work applies Fermi liquid theory to the spin current dynamics $\partial_{i}\vec{J}_{i}$ of ferromagnets.  
Given by Eq.~\eqref{dJ/dt}, its parameters are determined by the microscopic parameters obtained in Eqs.~\eqref{Pi-ik3} and \eqref{precessf}.  The same dynamical equation was obtained using Onsager's irreversible thermodynamics but with unknown parameters. In the paramagnetic limit, the spin current dynamics of ferromagnets agrees with that of paramagnets.

These results confirm the general structure found using Onsager's irreversible thermodynamics.\cite{SunSaslow24}  Moreover, they 
indicate that study of spin currents in ferromagnets can yield information about the Fermi liquid coefficients, 
as seen in Eqs. \eqref{GFL} and \eqref{lambdaFL}. The Appendix evaluates the magnetic susceptibility for a ferromagnet using Fermi liquid theory, thus providing an additional constraint of the parameters of that theory.

\section*{Acknowledgements}
\label{s:acknow}

We gratefully acknowledge the contributions of Shenglong Xu in the initial stages of this work. We also would like to thank Mark Stiles for introducing us to the problem of transport in conducting ferromagnets.  We gratefully acknowledge correspondence with C. J. Pethick.  C. S. was supported by the National Natural Science Foundation of China under No. 12105094, and by the Fundamental Research Funds for the Central Universities from China.

\appendix

\section{Fermi Liquid Interactions -- Longitudinal Susceptibility $\chi$}


Consider two Fermi surfaces with different Fermi wavevectors $k_{\uparrow,\downarrow}$ and momenta $p_{\uparrow,\downarrow}=\hbar k_{\uparrow,\downarrow}$ determined by the kinetic energies on the Fermi surface $\epsilon_{F,\uparrow,\downarrow}$ for each spin species.  The size difference between the two Fermi surfaces, due to exchange between the two Fermi surfaces, gives a net magnetization $M_{0}$ in equilibrium, which we take to be along $\hat{z}$.  It is given by
\bea
M_0 =-\frac{\gamma \hbar}{2}\int d\tau (n_\uparrow -n_\downarrow).
\label{M0}
\eea

We now calculate the longitudinal magnetic susceptibility $\chi$ by applying a small field $H=\delta H$ along $M_0$, to given a small $\delta M$.  We write
\begin{equation}
\delta n_\sigma(p) =\partial_\epsilon n_\sigma (\delta \epsilon_\sigma(p)-\delta \mu),
\label{dndeps}
\end{equation}
which includes a shift in chemical potential $\delta \mu$, determined by maintaining total particle number, so
\bea
0=\delta n_\uparrow +\delta n_\downarrow \equiv\int \delta n_\uparrow (p) d\tau + \int \delta n_\uparrow (p) d\tau .
\label{delN=0}
\eea

We also introduce the densities of states
\begin{equation}
{N_\uparrow}\equiv -\frac{dn_{\uparrow}}{d\epsilon}, \quad {N_\downarrow}\equiv -\frac{dn_{\downarrow}}{d\epsilon}.
\label{Ntildes}
\end{equation}
and write, with $\sigma=\pm1$ for $\uparrow\downarrow$,
\begin{equation}
\delta\epsilon_{\sigma}=\frac{\mu_0 \gamma \hbar }{2}\sigma g_{\sigma}H,
\label{deleps}
\end{equation}
where for the non-interacting case $g_{\sigma}=1$.

Including Fermi liquid interactions the energy changes due to $H$  are
\begin{eqnarray}
\delta \epsilon_\uparrow (p) &=& +\frac{\gamma\hbar\mu_0}{2} H + \int f_{\uparrow \sigma'} (p,p')\delta n_{\sigma'}(p') d\tau', \label{delepsintup} \\
\delta \epsilon_\downarrow (p) &=& - \frac{\gamma\hbar\mu_0}{2} H + \int f_{\downarrow \sigma'} (p,p')\delta n_{\sigma'}(p') d\tau'.
\label{delepsintdown}
\end{eqnarray}
Using \eqref{dndeps} and \eqref{deleps} 
in \eqref{delN=0} gives, on defining the dimensionless $\delta\tilde\mu$,
\begin{eqnarray}
\delta \mu \equiv \frac{\mu_0\gamma \hbar }{2} \,\frac{g_\uparrow \tilde{N_\uparrow} - g_\downarrow \tilde{N_\downarrow}}{ \tilde{N_\uparrow} +\tilde{N_\downarrow}} \, H \equiv \delta \tilde\mu \frac{\mu_{0}\gamma\hbar}{2}.
\label{delmuint}
\end{eqnarray}

We now determine $g_{\uparrow}$ and $g_{\downarrow}$ from \eqref{deleps} by applying \eqref{delepsintup} and \eqref{delepsintdown}.  We will retain only the uniform interactions $f^{(s)}_{0}$ and $f^{(a)}_{0}$,  in terms of which the $f_{\sigma \sigma'}$ read:
\begin{eqnarray}
f_{\uparrow\uparrow}=f_{\downarrow\downarrow}=  f^{(s)}_{0}+f^{(a)}_{0}, \quad 
f_{\uparrow\downarrow}=f_{\downarrow \uparrow}= f^{(s)}_{0}-f^{(a)}_{0}.
\label{fsfa}
\end{eqnarray}
We find that
\begin{eqnarray}
\delta \epsilon_\uparrow (p)
&=& +\frac{\gamma\hbar\mu_0}{2} H + {N}_{\uparrow}f_{\uparrow\uparrow}(\delta \mu - \frac{\mu_0 \gamma \hbar }{2} g_{\uparrow}H) \cr
&&+ {N}_{\downarrow}f_{\uparrow\downarrow}(\delta \mu + \frac{\mu_0 \gamma \hbar }{2} g_{\downarrow}H) ,
\label{delepsint2up} \\
\delta \epsilon_\downarrow (p)
&=& - \frac{\gamma\hbar\mu_0}{2} H + {N}_{\uparrow}f_{\downarrow\uparrow}(\delta \mu - \frac{\mu_0 \gamma \hbar }{2} g_{\uparrow}H) \cr
&&+ {N}_{\downarrow}f_{\downarrow\downarrow}(\delta \mu + \frac{\mu_0 \gamma \hbar }{2} g_{\downarrow}H).
\label{delepsint2down}
\end{eqnarray}



From \eqref{deleps}, \eqref{delepsint2up}, and \eqref{delepsint2down}, the self-consistent equations for $g_\uparrow$ and $g_\downarrow$ are
\begin{eqnarray}
g_\uparrow &=& 1 - {N}_{\uparrow}f_{\uparrow\uparrow}(g_\uparrow-\delta\tilde{\mu}) + {N}_{\downarrow}f_{\uparrow\downarrow}(g_\downarrow +\delta\tilde{\mu}), \quad\qquad\\
g_\downarrow &=& 1+ {N}_{\uparrow}f_{\downarrow\uparrow}(g_\uparrow-\delta\tilde{\mu}) - {N}_{\downarrow}f_{\downarrow\downarrow}(g_\downarrow +\delta\tilde{\mu}).
\label{gs}
\end{eqnarray}

Using Eqs.~\eqref{fsfa}, \eqref{delmuint}, and the definitions
\begin{equation}
F_{\sigma}^{(s,a)}={N}_{\sigma}f_{0}^{(s,a)},
\label{Fs}
\end{equation}
we solve for $g_\uparrow$ and $g_\downarrow$:
\begin{eqnarray}
g_\uparrow 
&=&\frac{{N_\downarrow} (1-2F_{\uparrow}^{(s)}) +{N_\uparrow}(1+2F_{ \downarrow}^{(s)})}{{N_\downarrow} (1+2F_{\uparrow}^{(a)}) +{N_\uparrow}(1+2F_{ \downarrow}^{(a)})},
\label{g2up} \\
g_\downarrow
&=&\frac{{N_\downarrow} (1+2F_{\uparrow}^{(s)}) +{N_\uparrow}(1-2F_{ \downarrow}^{(s)})}{{N_\downarrow} (1+2F_{\uparrow}^{(a)}) +{N_\uparrow}(1+2F_{ \downarrow}^{(a)})}.
\label{g2down}
\end{eqnarray}
\\

The longitudinal susceptibility is
\begin{eqnarray}
\chi & = &- \frac{\gamma \hbar}{2 \delta H} \int (\delta n_\uparrow(p) -\delta n_\downarrow(p) ) d\tau \nn\\
&=& \mu_0\frac{( \gamma \hbar)^2}{4}  \left[(g_\uparrow {N_\uparrow} + g_\downarrow {N_\downarrow }) -\delta \tilde{\mu} ({N_\uparrow} - {N_\downarrow} ) \right ] \nn\\
&=& \mu_0\frac{( \gamma \hbar)^2}{2} \frac{{N_\uparrow} {N_\downarrow}}{ {N_\uparrow} + {N_\downarrow }}(g_\uparrow + g_\downarrow )\nn\\
&=&\frac{\mu_0  (\gamma \hbar)^2 }{  (1+2F_{\uparrow}^{(a)} ){N_\uparrow} ^{-1} +(1+2 F_{\downarrow}^{(a)}){N_\downarrow} ^{-1} }.
\end{eqnarray}
Substituting from \eqref{Fs} gives
\bea
\chi =  \frac{\mu_0(\gamma \hbar)^2 }{({N_\uparrow} ^{-1}+{N_\downarrow} ^{-1})
+4f^{(a)}_{0}  }.
\label{chipar}
\eea
Thus, $\chi$ depends on $f^{(a)}_{0}$ but not on $f^{(s)}_{0}$. This result agrees with the paramagnetic limit.\cite{note-chi}


\begin{thebibliography}{11}

\bibitem{Wolf05} S. A. Wolf, D. D. Awschalom, R. A. Buhrman, J. M. Daughton, S. von Molnar, M. L. Roukes, A. Y. Chtchelkanova, and D. M. Treger, Science 294, 1488 (2005), ``Spintronics: a spin-based electronics vision for the future.''

\bibitem{ZFD-RMP} I. \v{Z}uti\'{c}, J. Fabian, and S. Das Sarma, Rev. Mod. Phys. 76, 323 (2004), ``Spintronics: Fundamentals and applications.''

\bibitem{Chumak15} A. V. Chumak, V. I. Vasyuchka, A. A. Serga and B. Hillebrands,  Nat. Phys. 11, 453 (2015), ``Magnon spintronics.''

\bibitem{Hirohata20} A. Hirohata, K. Yamada, Y. Nakatani, I.-L. Prejbeanu, B. Di\'{e}ny, P. Pirro, and B. Hillebrands, J. Magn. Magn. Mater. 509, 166711 (2020), ``Review on spintronics: Principles and device applications.''

\bibitem{Kim24} K.-W. Kim, B-G. Park, and K.-J. Lee, NPJ Spintronics 2, 8 (2024), ``Spin current and spin-orbit torque induced by ferromagnets.''


 
\bibitem{JohnsonSilsbee87} M. Johnson and R. H. Silsbee, Phys. Rev. B 35, 4959 (1987), ``Thermodynamic analysis of interfacial transport and of the thermomagnetoelectric system.''

\bibitem{Saslow07} W. M. Saslow, Phys. Rev. B 76, 184434 (2007), ``Spin pumping of current in non-uniform conducting magnets.''

\bibitem{Saslow15} W. M. Saslow, Phys. Rev. B 91, 014401 (2015), ``Spin Hall effect and irreversible thermodynamics: Center-to-edge transverse current-induced voltage.''

\bibitem{Saslow17} W. M. Saslow, Phys. Rev. B 95, 184407 (2017), ``Irreversible thermodynamics of uniform ferromagnets with spin accumulation: Bulk and interface dynamics.''

\bibitem{SaslowSunXu22} W. M. Saslow, C. Sun, and S. Xu, Phys. Rev. B 105, 174441 (2022), ``Spin accumulation and longitudinal spin diffusion of magnets.''


\bibitem{SunSaslow23} C. Sun and W. M. Saslow, Phys. Rev. B 107, 094438 (2023), ``Spin diffusion in spin glasses requires two magnetic variables, $\vec{M}$ and $\vec{m}$.''



\bibitem{SunSaslow24} C. Sun and W. M. Saslow, arXiv:2402.04639 [cond-mat.mes-hall] (7 Feb 2024), ``Macroscopic Magnetic Dynamics.''

\bibitem{Leggett70} A. J. Leggett, J. Phys. C 3, 448-459 (1970), ``Spin diffusion and spin echoes in liquid $^{3}$He at low temperature.''

\bibitem{LandauFLT56} L. D. Landau, Zh. Eksp. i Teor. Fiz. 30, 1058 (1956), Soviet Phys. JETP  3, 920 (1957), ``The Theory of a Fermi Liquid.''

\bibitem{LandauFLT57} L. D. Landau, Zh. Eksp. Teor. Fiz. 32, 59 (1957) [Sov. Phys. JETP 5, 101 (1957)], ``Oscillations in a Fermi Liquid.''

\bibitem{Silin2} V. P. Silin, Zh. Eksp. i Teor. Fiz. 33, 495 (1957), Soviet Phys. JETP 6, 387 (1958), ``Theory of a degenerate Fermi liquid.''

\bibitem{Silin1} V. P. Silin, Zh. Eksp. i Teor. Fiz. 33, 127 (1957), Soviet Phys. JETP 6, 945 (1958), ``Oscillations of a Fermi liquid in a magnetic field.''


\bibitem{Mineev04} V. P. Mineev, Phys. Rev. B, 69, 144429 (2004), ``Transverse spin dynamics in a spin-polarized Fermi liquid.''

\bibitem{Mineev05} V. P. Mineev, Phys. Rev. B, 72, 144418 (2005),  ``Theory of transverse spin dynamics in a polarized Fermi liquid and an itinerant ferromagnet''.

\bibitem{Bedell86} K. S. Bedell and C. Sanchez-Castro, Phys. Rev. Lett. 57, 854 (1986), ``Near-Metamagnetism of Liquid $^3$He at High Pressure.''

\bibitem{JeonMullin89} J. W. Jeon and W. J. Mullin, Phys. Rev. Lett. 62, 2691 (1989), ``Transverse spin diffusion in polarized Fermi gases.''

\bibitem{MiyakeMullinStamp85} K. Miyake, W. J. Mullin, and P. C. E. Stamp, J. Physique 46, 663-671 (1985),  ``Mean-field and spin-rotation phenomena in Fermi systems: the relation between the Leggett-Rice and Lhuillier-Lalo\"e effects.''

\bibitem{GolosovRuckenstein95}  D. I. Golosov and A. E. Ruckenstein, Phys. Rev. Lett. 74, 1613 (1995), ``Low-Temperature Spin Diffusion in a Spin-Polarized Fermi Gas.''

\bibitem{Levitov02} M. \"O. Oktel and L. S. Levitov, Phys. Rev. Lett. 88, 230403 (2002), ``Internal Waves and Synchronized Precession in a Cold Vapor.''



\bibitem{BaymPethick78} G. Baym and C. Pethick, Chapter 1 of The Physics of Liquid and Solid Helium, Part II, edited by K. H. Bennemann and J. B. Ketterson (Wiley, New York, 1978).

\bibitem{Pethickprivate} This term is double that of BP (1.3.89), which may be a misprint. C. J. Pethick informs us that he has confirmed our result in the paramagnetic limit.

\bibitem{note-chi}
In the paramagnetic limit $\tilde{N_\uparrow}=\tilde{N_\downarrow}=N(0)/2$, and $\chi = \mu_0(\gamma \hbar)^2 N(0)/[4 (1+  N(0)f^{(a)}_{0} )]$, which agrees with BP's (1.1.55).







\end{thebibliography}
\end{document}